\documentclass[prd,aps,twocolumn,nofootinbib,floatfix,10pt]{revtex4}
\usepackage{graphicx,epsfig,mathtools}
\usepackage[usenames]{color}
\usepackage{slashed}
\usepackage{epstopdf}
\usepackage{multirow}

\begin{document}
\title{Baryonic heavy-to-light form factors induced by   tensor current in light-front approach}

\author{Hang Liu$^1$, Wei Wang$^1$~\footnote{Email:wei.wang@sjtu.edu.cn}, Zhi-Peng Xing$^2$~\footnote{Email:zpxing@sjtu.edu.cn} }

\affiliation{$^{1}$ INPAC, Key Laboratory for Particle Astrophysics and Cosmology (MOE), Shanghai Key Laboratory for Particle Physics and Cosmology,
School of Physics and Astronomy, Shanghai Jiao Tong University, Shanghai 200240, China}
\affiliation{$^{2}$  Tsung-Dao Lee Institute, Shanghai Jiao Tong University, Shanghai 200240, China}

\begin{abstract}
Inspired by the recent  anomalies on the lepton flavor universalities, 
we present an investigation of baryonic form factors induced by  heavy-to-light tensor currents.  With the light-front quark model, we calculate the tensor form factors, and the momentum distributions are accessed with two parametrization forms. Numerical results for the form factors at $q^2=0$ are derived and parameters in $q^2$ distributions are obtained through extrapolation. Our results are helpful for the analysis of new physics contributions in the heavy-to-light transition. 
\end{abstract}

\maketitle

\section{Introduction}
Nowadays testing the standard model (SM) of particle physics and hunting for new physics (NP) beyond the SM is a foremost task in particle physics.  In recent years, heavy flavor physics has received remarkable  attentions  and quite interestingly some anomalies  are found in weak decays of heavy mesons and baryons.  Hints for the lepton flavor universality (LFU) deviating from the standard model are found recently in ratios of branching fractions such as $R(D^{(*)})$\cite{BaBar:2012obs,BaBar:2013mob,Belle:2015qfa,Belle:2016ure,Belle:2016dyj,Belle:2017ilt,Belle:2019rba}, $R(J/\psi)$\cite{LHCb:2017vlu},  and $R(\Lambda_c)$\cite{LHCb:2022piu}. A collection of the latest experimental measurement is given as~\cite{HeavyFlavorAveragingGroup:2022wzx}
\begin{eqnarray}
&&{\mathcal R}_{exp}(D)=0.358\pm0.025\pm0.012,\notag\\
&&{\mathcal R}_{exp}(D^*)=0.285\pm0.010\pm0.008,\notag\\
&&{\mathcal R}_{exp}(J/\psi)=0.71\pm0.17\pm0.18,\notag\\
&&{\mathcal R}_{exp}(\Lambda_c)=0.242\pm0.026\pm0.040\pm0.059,
\end{eqnarray}
while the recent average of SM predictions on $R_{D/D^*}$ is given as~\cite{HeavyFlavorAveragingGroup:2022wzx}: 
\begin{eqnarray}
&&{\mathcal R}_{SM}(D)=0.298\pm0.004,\notag\\
&&{\mathcal R}_{SM}(D^*)=0.254\pm0.005.
\end{eqnarray}
One can see that the latest experimental measurement on $R(D^{(*)})$ shows about a $3\sigma$ standard deviation from SM prediction. While the most precise SM predictions on $R(J/\psi)$ and $R(\Lambda_c)$  are generally consistent with experimental measurements~\cite{Harrison:2020nrv,Bernlochner:2018bfn,Iguro:2022yzr}:
\begin{eqnarray} 
&&{\mathcal R}_{SM}(J/\psi)=0.258\pm 0.004,\notag\\
&&{\mathcal R}_{SM}(\Lambda_c)=0.324\pm0.004.
\end{eqnarray}
there are also noticeable discrepancies in the central values between the SM predictions and experimental results. Thereby,  more theoretical and experimental investigations are needed to clarify this  obscure situation. 


Inspired by these deviations,  many interesting theoretical explanations have emerged to include the new physics contributions, and for a recent review, please see Ref.~\cite{Li:2018lxi}. In  some of these analyses new   effective  NP Hamiltonian have been introduced~\cite{Shivashankara:2015cta,Li:2016pdv,Hu:2018luj,Zhao:2020mod,Shen:2021dat,Tao:2022yur,Biancofiore:2013ki,Tang:2022nqm,Du:2022ipt,Fedele:2022iib}. Unlike the SM, some of these Hamiltonians introduce a new structure of $b\to c$ current in the hadron matrix element: the tensor current. 
While the tensor current contribution resolves the current discrepancies,  it should be noticed that this contribution can also affect baryonic decay processes  induced by the $b\to c\ell\nu$. Though the explicit contributions of the tensor current depend on the  details  of NP models,  the  low energy nonperturbative matrix elements, namely form factors, are universal and mandatory. Therefore for a comprehensive NP model-independent analysis,  the results on baryonic tensor form factors  are  called for.


In this work,  we aim to present an explortion  of these new form factors, and in the calculation  employ the light-front quark model (LFQM) ~\cite{Jaus:1989au,Jaus:1991cy,Cheng:1996if,Jaus:1999zv,Cheng:2003sm,Scadron:2003yg}. The baryon system   can be treated in a similar  manner with  a meson under  the quark-diquark assumption~\cite{Xing:2018lre,Zhao:2018mrg,Zhao:2022vfr}. It is noteworthy that a three-quark vertex function has been explored in LFQM, and the  investigations in Ref.~\cite{Ke:2019smy} and Ref.~\cite{Lu:2023rmq} have shown consistent results with the two approaches using a three-quark and diquark vertex. This also validates the diquark approach that will be adopted  in this work.

The rest of this paper is arranged as follows. We give the definition of tensor form factors and
the theoretical framework of LFQM in Section II.
The tensor form factors are explicitly calculated in Section III, and the analytical expressions are given. 
Numerical analysis and discussions are given in Section IV. In
the end we give a brief summary.

\section{Theoretical  framework}

To account for the anomalies in heavy quark physics, NP contributions are typically needed.  In a general analysis, the Hamiltonian after integrating out the degree of freedoms over the $m_b$ scale can be expressed  as
\begin{eqnarray}
{\mathcal H}_{eff}\sim \sum_{\Gamma,\Gamma^\prime}C_{\Gamma,\Gamma^\prime}\times [\bar c \Gamma b][ \bar l\Gamma^\prime \nu],
\end{eqnarray}
where the $C_{\Gamma,\Gamma^\prime}$  are the Wilson coefficients and the $\Gamma^{(\prime)}$ denote all the possible Lorentz structure $\{ 1,\gamma^\mu,\gamma_5,\gamma^\mu\gamma_5,\sigma^{\mu\nu}\}$.  The tensor current does not exist  in the SM, and as a result the corresponding  form factors are not well understood in previous theoretical analyses~\cite{Huang:2018nnq,Shen:2021dat,Tao:2022yur,Tang:2022nqm,Harrison:2023dzh}. 

This work will focus on the anti-triplet b-baryon to anti-triplet c-baryon and sextet b-baryon to sextet c-baryon processes: 
\begin{eqnarray}
&&\bar 3\to \bar 3: \;\;\;\Lambda_b^0\to\Lambda_c^+,\enspace\Xi_b^0\to\Xi_c^+,\enspace\Xi_b^-\to\Xi_c^0\notag\\
&&6\to 6: \;\;\; \Sigma_b^+\to\Sigma_c^{++},\enspace\Sigma_b^-\to\Sigma_c^0,\enspace\Sigma_b^0\to\Sigma_c^+,\notag\\
&&\;\;\;\;\;\;\;\;\;\;\;\;\;\;\enspace\Xi_b^{\prime 0}\to\Xi_c^{\prime +},\enspace\Xi_b^{\prime -}\to\Xi_c^{\prime 0},\enspace\Omega_b^-\to\Omega_c^0.
\end{eqnarray}
Inserting the Dirac Gamma matrices in bilinear local
quark current sandwiched between the b-baryons and c-baryons, one can define  the pertinent form factors: 
\begin{eqnarray}
&&\langle {\mathcal B}^\prime(P^\prime,S^\prime)|\bar c i\sigma^{\mu\nu}b|{\mathcal{B}(P,S)}\rangle\notag\\
&&=\bar u(P^\prime,S^\prime)\bigg[
f_1(q^2)i\sigma^{\mu\nu}+\Sigma^{\mu\nu}_{\sigma\rho}(\frac{f_2(q^2)}{M} \gamma^\sigma P^\rho\notag\\
&&\;\;\;+\frac{f_3(q^2)}{M^\prime} \gamma^\sigma P^{\prime\rho}+\frac{f_4(q^2)}{MM^\prime}P^\sigma P^{\prime\rho})\bigg]u(P,S),\label{eq:paremetrization}
\end{eqnarray}
where Dirac operator $\sigma^{\mu\nu}=\frac{i}{2}(\gamma^{\mu}\gamma^{\nu}-\gamma^{\nu}\gamma^{\mu})$ and $\Sigma^{\mu\nu}_{\sigma\rho}=g^\mu_\sigma g^\nu_\rho-g^\nu_\sigma g^\mu_\rho$. The form factors in the $\langle {\mathcal B}^\prime(P^\prime,S^\prime)|\bar c\sigma^{\mu\nu}\gamma_5b|{\mathcal{B}(P,S)}\rangle$ matrix element can be straightforwardly obtained from Eq.~\eqref{eq:paremetrization} by making use of $\sigma ^{\mu\nu}\gamma_5=-\frac{i}{2}\epsilon^{\mu\nu\lambda\rho}\sigma_{\lambda\rho}$.

The form factors defined in hadronic matrix element are  non-perturbative. To evalulate the form factors a relativistic quark model, namely light-front quark model, will be employed.  In this framework, it is convenient to use the light-front frame: 
\begin{eqnarray}
p^\mu=(p^+,p^-,p_\perp),\quad p^+=p^0+p^3,\nonumber\\
p^-=p^0-p^3,\quad p_\perp=(p^1,p^2).
\end{eqnarray}
With the help of the diquark picture, the baryon state containing three quarks can be treated as a meson state which is wide studied in the LFQM~\cite{Hu:2020mxk}. In this picture, the two spectator quark played a simiarl role with  an anti-quark.

For the $J^P=1/2^+$ baryon state, its wave function is
\begin{eqnarray}
	&& |{\cal B}(P,S,S_{z})\rangle  =  \int\{d^{3}p_{1}\}\{d^{3}p_{2}\}2(2\pi)^{3}\delta^{3}(\tilde{P}-\tilde{p}_{1}-\tilde{p}_{2})\nonumber \\
	&  & \times\sum_{\lambda_{1},\lambda_{2}}\Psi^{SS_{z}}(\tilde{p}_{1},\tilde{p}_{2},\lambda_{1},\lambda_{2})|Q(p_{1},\lambda_{1})({\rm{di}})(p_{2},\lambda_{2})\rangle,\label{eq:state_vector}
\end{eqnarray}
where $Q$ donates the heavy quark which is $b/c$ in our work and  ``$({\rm{di}})$" presents the diquark.  The $\lambda_1$ and $\lambda_2$ are the helicity of quark and diquark respectively.
The momentum of them are $p_1$ and $p_2$ and the $P$ is the baryon momentum.  Both of them are on their mass shell. However, since the baryon states are constructed by quark and diquark, their momenta can not be on shell simultaneously.
Thus in the wave function of LFQM the three momentum $\{\tilde P, \tilde p_1, \tilde p_2\}$ are used. They are defined as $\tilde{p}=(p^+,p_{\perp})$. The distribution function  $\Psi$ in Eq.~\eqref{eq:state_vector} is
\begin{eqnarray}
	\Psi^{SS_{z}}(\tilde{p}_{1},\tilde{p}_{2},\lambda_{1},\lambda_{2})=\frac{1}{\sqrt{2(p_{1}\cdot\bar{P}+m_{1}M_{0})}} \nonumber\\
 \times \bar{u}(p_{1},\lambda_{1})\Gamma_{S(A)} u(\bar{P},S_{z})\phi(x,k_{\perp}).\label{eq:momentum_wave_function_1/2}
\end{eqnarray}
Here $\Gamma$ is the coupling vertex which embody the spin of diquark. For the spin-0 scalar diquark the vertex is $\Gamma_S$ and for the spin-1 axis-vector diquark the vertex becomes $\Gamma_A$. The coupling vertices are shown as
\begin{eqnarray}
&&	\Gamma_S=1,\nonumber\\
&& \Gamma_{A} =\frac{\gamma_{5}}{\sqrt{3}}\left(\slashed\epsilon^{*}-\frac{M_0+m_1+m_2}{\bar{P}\cdot p_2+m_2M_0}\epsilon^{*}\cdot\bar{P}\right),\label{eq:momentum_wave_function_1/2gamma}
\end{eqnarray}
where the $\bar P$ is the sum of on shell momenta $p_1$ and $p_2$. Though momentum $\bar P$ can not on its mass shell, one can defined $M_0^2=\bar P^2$ with $M^2\neq M_0^2$.

The $\phi$ in Eq.~\eqref{eq:momentum_wave_function_1/2gamma} is a Gaussian-type momentum distribution function which is constructed as
\begin{equation}
	\phi=4\left(\frac{\pi}{\beta^{2}}\right)^{3/4}\sqrt{\frac{e_{1}e_{2}}{x_{1}x_{2}M_{0}}}\exp\left(\frac{-\vec{k}^{2}}{2\beta^{2}}\right),\label{eq:Gauss}
\end{equation}
where $e_1$ and $e_2$ represent the energy of  heavy quark $Q$ and diquark in the rest frame of $\bar{P}$.
The $\beta$ is phenomenological parameter which is shown in Table.~\ref{table1}.   In Ref.~\cite{Li:2022hcn}, a Gaussian expansion method with a semirelativistic potential model is applied to determine the momentum distribution wave function and in Ref.~\cite{Zhao:2023yuk}, the parameters $\beta$ are extracted by the pole residue. In this work the parameters $\beta$ are used from  Ref.~\cite{Hu:2020mxk}.

The $x_1$ and $x_2$ in Eq.~\eqref{eq:Gauss} are the light-front momentum fractions which satisfy the requirements  $0<x_2<1$ and $x_1+x_2=1$. The $k$ is the  internal momentum which represent the interaction between quark and diquark. So the $k$ and the quark and diquark momenta are
\begin{eqnarray}
	 &&k_i=(k_i^-,k_i^+,k_{i\bot})=(e_i-k_{iz},e_i+k_{iz},k_{i\bot}) \nonumber\\
  && =
  (\frac{m_i^2+k_{i\bot}^2}{x_iM_0},x_iM_0,k_{i\bot}),\nonumber\\
	&&p^+_1=x_1\bar P^+,   ~~~p^+_2=x_2 \bar P^+,  ~~~p_{1\perp}=x_1 \bar P_{\perp}+k_{1\perp},\nonumber\\
  &&   p_{2\perp}=x_2 \bar P_{\perp}+k_{2\perp},
  ~~~ k_{\perp}=-k_{1\perp}=k_{2\perp}.\label{eq:im}
\end{eqnarray}
With the help of the internal momentum $k$, the invariant mass square $M_0^2$ is expressed as
\begin{eqnarray} \label{eq:Mpz}
	  M_0^2=\frac{k_{1\perp}^2+m_1^2}{x_1}+ \frac{k_{2\perp}^2+m_2^2}{x_2}.
 \end{eqnarray}
 The expression of $e_i$ and $k_z$ can also be presented in terms of the internal variables $(x_{i},k_{i\bot})$ as
\begin{eqnarray}
&& e_i=\frac{x_iM_0}{2}+\frac{m_i^2+k_{i\perp}^2}{2x_iM_0} =\sqrt{m_i^2+k_{i\bot}^2+k_{iz}^2}, \nonumber\\
&&  k_{iz}=\frac{x_iM_0}{2}-\frac{m_i^2+k_{i\perp}^2}{2x_iM_0}.
 \end{eqnarray}
In the following,  we use the notation $x=x_2$ and  $x_1=1-x$. 

\section{Form factors}
The hadronic matrix element in Eq.~\eqref{eq:paremetrization}  can be expressed by the LFQM as
\begin{eqnarray}
&&	 \langle{\cal B}_{c}(P^{\prime}, \frac{1}{2},S_{z}^{\prime})|\bar{c}i\sigma^{\mu\nu}b|{\cal B}_{b}(P, \frac{1}{2},S_{z})\rangle\nonumber\\
  && =  \int\{d^{3}p_{2}\}\frac{\phi^{\prime}(x^{\prime},k_{\perp}^{\prime})\phi(x,k_{\perp})}{2\sqrt{p_{1}^{+}p_{1}^{\prime+}(p_{1}\cdot\bar{P}+m_{1}M_{0})(p_{1}^{\prime}\cdot\bar{P}^{\prime}+m_{1}^\prime M_{0}^{\prime})}}\notag\\
 &&\times\sum_{\lambda_{2}}\bar{u}(\bar{P}^{\prime},S_{z}^{\prime})\left[\bar{\Gamma}^{\prime}(\slashed p_{1}^{\prime}+m_{c})\sigma^{\mu\nu}(\slashed p_{1}+m_{b})\Gamma\right] u(\bar{P},S_{z}), \nonumber\\\label{eq:matrix_element_onehalf}
\end{eqnarray}
with
\begin{eqnarray}
\bar{P}^\prime=p_{1}^{\prime}+p_{2},\quad M_{0}^2=\bar{P}^{2},\quad M_{0}^{\prime2}=\bar{P}^{\prime2}.\label{parameter}
\end{eqnarray}
Since the matrix element can be expressed both at the quark and haron levels respectively, the form factors can be extracted by solving eight equations which are constructed by multiplying the different  scalar $ \bar u(P,S)\{\Gamma_i\}_{\mu\nu} u(P^\prime,S^\prime)$ to the matrix element in light-front approach. The Lorentz structure in these  scalars are 
\begin{eqnarray}
\{\Gamma_i\}_{\mu\nu}=\{\gamma_\mu P^\prime_\nu,\gamma_\mu P_\nu,P_\mu P^\prime_\nu,\gamma_\mu\gamma_\nu\}.
\end{eqnarray}
Then these form factors are  calculated as%
\begin{eqnarray}
f_1&=&\frac{1}{4s_+s_-}[4 M H_1-4 M^\prime H_2-4H_3+H_4 s_-],\notag\\
f_2&=&-\frac{1}{2s_-^2s_+}[M(s_++2MM^\prime)H_1\notag\\
&&-6MM^{\prime2}H_2-6MM^{\prime}H_3+MM^\prime H_4s_-],\notag\\
f_3&=&\frac{1}{2s_-^2s_+}[6M^2 M^\prime H_1-M^\prime(s_++2MM^\prime)H_2\notag\\
&&-6MM^\prime H_3+MM^\prime H_4s_-],\notag\\
f_4&=&-\frac{1}{2s_-^2s_+^2}[6M^2 M^\prime H_1s_+-6MM^{\prime2} H_2s_+\notag\\
&&-24M^2M^{\prime2} H_3+MM^\prime H_4s_-s_+],
\end{eqnarray}
where the function $H_i$ is defined as
\begin{eqnarray} 
	H_i&=&\int\frac{dxd^2k_\perp}{2(2\pi)^3}\phi^\prime(x^\prime,k^\prime_\perp)\phi(x,k_\perp)\times {\rm Tr}\big [(\slashed{\bar{P}}^\prime+M^\prime_0)\Gamma^\prime_{S(A)}\notag\\
 &&\times (\slashed p^\prime_1+m_c)\sigma_{\mu\nu}  (\slashed p_1+m_b)\Gamma_{S(A)}(\slashed{\bar{P}}+M_0)\{\Gamma_i\}_{\mu\nu}\big ]\notag\\
 &&\times\bigg(2\sqrt{x^\prime_1x_1(p^\prime_1\cdot\bar{P}^\prime+m^\prime_1M^\prime_0)(p_1\cdot\bar{P}+m_1M_0)}\bigg)^{-1}\notag\\
\end{eqnarray}

In our analysis with the diquark picture, the diquark made of two quarks can be  a scalar or an axial-vector. Thus the physical form factors estimated by the LFQM should be expressed as the combination of the form factors with scalar and axis vector diquark:
 \begin{eqnarray}
&& F^{[phy]}=c_S \times  F_S + c_A\times  F_A.
 \end{eqnarray}
 The coefficients $c_S$ and $c_A$ can be extracted in the flavor-spin wave function of initial and final baryon states.

 Using the scalar and axial-vector diquark wave function~\cite{Wang:2022ias}: $[q,q^\prime]_A=(qq^\prime+q^\prime q)/\sqrt{2}$, $[q,q^\prime]_S=(qq^\prime-q^\prime q)/\sqrt{2}$, one can write the flavor-spin wave function of baryon. Since the baryons $\{\Lambda_b^0,\Xi_b^0,\Xi_b^-\}$ and $\{\Lambda_c^+,\Xi_c^+,\Xi_c^0\}$ are the anti-triplet in $SU(3)$ flavor symmetry, their flavor-spin wave function are
 \begin{eqnarray}
&&|{\cal B}_{Qqq^\prime},S_z=\frac{1}{2}\rangle=\frac{1}{\sqrt{2}}(qq^\prime-q^\prime q)Q\times \bigg(\frac{1}{\sqrt{2}}(\uparrow \downarrow\uparrow-\downarrow\uparrow\uparrow)\bigg) \nonumber\\
&&\equiv [q,q^\prime]_SQ.
\end{eqnarray}
For the sextet baryons, their flavor-spin wave function can be written as
\begin{eqnarray}
&& |{\cal B}_{Qqq^\prime},S_z=\frac{1}{2}\rangle=\frac{1}{\sqrt{2}}(qq^\prime+q^\prime q)Q \nonumber\\
&&\times \bigg(\frac{1}{\sqrt{6}}(\uparrow \downarrow\uparrow+\downarrow\uparrow\uparrow-2\uparrow \uparrow\downarrow)\bigg)\equiv -[q,q^\prime]_AQ,\notag\\
&&|{\cal B}_{Qqq},S_z=\frac{1}{2}\rangle=qqQ\times \bigg(\frac{1}{\sqrt{6}}(\uparrow \downarrow\uparrow+\downarrow\uparrow\uparrow-2\uparrow \uparrow\downarrow)\bigg)\nonumber\\
&&\equiv -[q,q]_AQ.
\end{eqnarray}
where $q,q'=u,d,s$ and $Q=b,c$. For a scalar diquark, namely the baryon triplet, $c_S=1$ and $c_A=0$.  For the sextet baryons with an axial-vector diquark,  $c_S=0$  and $c_A=1$.
It is worth noting that several studies have suggested that the flavor eigenstates $\Xi_c$ and $\Xi^\prime_c$ may mix with each other to generate the mass eigenstates~\cite{Aliev:2010ra,Matsui:2020wcc,He:2021qnc,Geng:2022yxb,Ke:2022gxm,Xing:2022phq}. But the recent  Lattice QCD indicates  a very small mixing angle~\cite{Liu:2023feb} which is consistent with previous Lattice QCD simulation~\cite{Brown:2014ena}. As a consequence  the mixing effect is not taken into account in our analysis.
\section{Numerical analysis}

Numerical results of tensor form factors will be given in this section. In our calculation, the masses and other parameters of these baryons are shown in Table.~\ref{table1}.

\begin{center}
 \begin{table}[!htb]
\caption{Masses of charm and bottom baryons and input parameters in the momentum distribution wavefunctions~\cite{Hu:2020mxk}. }
\label{table1}%
\begin{tabular}{|c|c|c|c|c|c|c|c|c|c|}
\hline \hline
baryons & $\Lambda_b^0$  & $\Sigma_b^+$ &$\Sigma_b^-$&$\Sigma_b^0$&$\Xi_b^0$&$\Xi_b^-$ \tabularnewline
\hline
mass$({\rm GeV})$ &5.620&5.811&5.816&5.814&5.792&5.797\tabularnewline
\hline
baryons&$\Xi_b^{\prime 0}$&$\Xi_b^{\prime -}$&$
\Omega_b^-$& $\Lambda_c^+$  & $\Sigma_c^{++}$ &$\Sigma_c^0$  \tabularnewline\hline
mass$({\rm GeV})$&5.935&5.935&6.045&2.286&2.454&2.454\tabularnewline
\hline
baryons &$\Sigma_c^+$&$\Xi_c^+$&$\Xi_c^0$&$\Xi_c^{\prime +}$&$\Xi_c^{\prime 0}$&$\Omega_c^0$  \tabularnewline
\hline
mass$({\rm GeV})$  &2.453&2.468&2.470&2.578&2.579&2.695\tabularnewline
\hline
\multirow{2}{*}{shape parameter}&\multicolumn{3}{|c|}{$\beta_{c[uq]}=0.470$}&\multicolumn{3}{|c|}{$ \beta_{c[sq]}=0.535$}
\cr\cline{2-7}
&\multicolumn{3}{|c|}{$ \beta_{b[uq]}=0.562$}&\multicolumn{3}{|c|}{$ \beta_{b[sq]}=0.623$}\tabularnewline
\hline\hline
\end{tabular}
\end{table}
\end{center}

For the calculation, we use the constituent  quark masses from Refs.~\cite{Li:2010bb,Verma:2011yw,Shi:2016gqt}: 
\begin{eqnarray}
	 && m_u=m_d= 0.25~{\rm GeV}, m_s=0.37~{\rm GeV}, \nonumber \\
      &&m_c=1.4~{\rm GeV}.
\label{eq:mass_quark}
\end{eqnarray}
The masses of the diquarks can be approximated as
\begin{eqnarray}
	 m_{[q q^\prime]}=m_q+m_{q^\prime}, \quad q/q^\prime=u,d,s.\label{eq:mass_diquark}
\end{eqnarray}

\begin{widetext}

 \begin{table}[!http]
\caption{
Numerical results for the tensor form factors of b-baryon to c-baryon transitions induced by $b\to c$ decays are presented. The parameters $\delta$ and $m_{\rm fit}$ are the results of fitting the pole model in Eq.\ref{fit}, and the $^*$ indicates that $m_{\rm fit}$ is fitted using Eq.\ref{fit1}. The $a_0$ and $a_1$ are the parameters in the BCL model in Eq.~\ref{BCL}. For the form factors with $q^2=0$, i.e., $F(0)$, we estimated their uncertainties caused by the parameters in LFQM, namely, $\beta_{b[qq]}$, $\beta_{c[qq]}$, and $m_{di}$, which are varied by $10\%$.}
\label{ffnr3} %
\begin{tabular}{|c|c|c|c|c|c|c|c|c|}
\hline \hline
\multirow{2}{*}{channel}&\multirow{2}{*}{form factor}&\multicolumn{3}{c|}{Pole model}&\multicolumn{2}{c|}{BCL model}\cr\cline{3-7}

&& F(0) &$\delta$ & $m_{\rm fit}$ & $a_0$ & $a_1$  \\
\hline
\multirow{4}{*}{$\Lambda_b^0\to\Lambda_c^+$}&$ f_1$&$0.649\pm0.019\pm0.079\pm0.014$&1.30&6.15&0.66&-0.37\cr\cline{2-7}
&$ f_2$&$-0.167\pm0.039\pm0.159\pm0.071$&5.12&4.07*&-0.10&-2.61\cr\cline{2-7}
&$ f_3$&$0.105\pm0.024\pm0.149\pm0.072$&2.80&3.17*&0.05&2.47\cr\cline{2-7}
&$ f_4$&$-0.042\pm0.037\pm0.144\pm0.065$&0.92&1.79*&0.03&-3.01\tabularnewline
\hline\hline
\multirow{4}{*}{$\Sigma_b^+\to\Sigma_c^{++}$}&$ f_1$&$-0.196\pm0.006\pm0.019\pm0.005$&0.78&5.68&-0.20&0.34\cr\cline{2-7}
&$ f_2$&$0.057\pm0.046\pm0.037\pm0.010$&2105&21.96*&0.05&0.22\cr\cline{2-7}
&$ f_3$&$-0.056\pm0.039\pm0.032\pm0.012$&16.19&6.83&-0.06&0.10\cr\cline{2-7}
&$ f_4$&$0.401\pm0.027\pm0.047\pm0.009$&0.80&4.49&0.46&-2.78\tabularnewline
\hline\hline
\multirow{4}{*}{$\Sigma_b^-\to\Sigma_c^0$}&$ f_1$&$-0.197\pm0.006\pm0.019\pm0.005$&0.72&5.63&-0.21&0.37\cr\cline{2-7}
&$ f_2$&$0.062\pm0.046\pm0.037\pm0.010$&38.34&8.57&0.06&0.02\cr\cline{2-7}
&$ f_3$&$-0.061\pm0.039\pm0.032\pm0.012$&4.77&5.30&-0.07&0.30\cr\cline{2-7}
&$ f_4$&$0.404\pm0.027\pm0.047\pm0.009$&0.71&4.43&0.47&-2.94\tabularnewline
\hline\hline
\multirow{4}{*}{$\Sigma_b^0\to\Sigma_c^+$}&$ f_1$&$-0.197\pm0.006\pm0.019\pm0.005$&0.73&5.64&-0.20&0.37\cr\cline{2-7}
&$ f_2$&$0.060\pm0.046\pm0.037\pm0.010$&73.49&9.94&0.06&0.07\cr\cline{2-7}
&$ f_3$&$-0.060\pm0.039\pm0.032\pm0.012$&6.01&5.55&-0.07&0.25\cr\cline{2-7}
&$ f_4$&$0.404\pm0.027\pm0.047\pm0.009$&0.73&4.44&0.47&-2.90\tabularnewline
\hline\hline
\multirow{4}{*}{$\Xi_b^0\to\Xi_c^+$}&$ f_1$&$0.651\pm0.022\pm0.085\pm0.021$&1.80&6.44&0.65&-0.04\cr\cline{2-7}
&$ f_2$&$-0.214\pm0.050\pm0.195\pm0.100$&4.86&3.99*&-0.13&-3.76\cr\cline{2-7}
&$ f_3$&$0.140\pm0.031\pm0.184\pm0.102$&2.91&3.21*&0.06&3.49\cr\cline{2-7}
&$ f_4$&$-0.069\pm0.048\pm0.179\pm0.093$&1.14&2.03*&0.02&-4.12\tabularnewline
\hline\hline
\multirow{4}{*}{$\Xi_b^-\to\Xi_c^0$}&$ f_1$&$0.650\pm0.022\pm0.084\pm0.021$&1.73&6.41&0.65&-0.07\cr\cline{2-7}
&$ f_2$&$-0.214\pm0.049\pm0.195\pm0.100$&5.21&4.10*&-0.14&-3.63\cr\cline{2-7}
&$ f_3$&$0.141\pm0.031\pm0.183\pm0.102$&3.10&3.30*&0.07&3.35\cr\cline{2-7}
&$ f_4$&$-0.070\pm0.048\pm0.179\pm0.093$&1.20&2.08*&0.02&-3.97\tabularnewline
\hline\hline
\multirow{4}{*}{$\Xi_b^{\prime 0}\to\Xi_c^{\prime+}$}&$ f_1$&$-0.203\pm0.006\pm0.021\pm0.005$&0.88&5.68&-0.21&0.38\cr\cline{2-7}
&$ f_2$&$0.070\pm0.058\pm0.046\pm0.012$&171.79&11.97&0.07&0.12\cr\cline{2-7}
&$ f_3$&$-0.066\pm0.049\pm0.040\pm0.015$&7.33&5.63&-0.07&-0.31\cr\cline{2-7}
&$ f_4$&$0.434\pm0.035\pm0.056\pm0.010$&0.86&4.52&0.50&-3.13\tabularnewline
\hline\hline
\multirow{4}{*}{$\Xi_b^{\prime -}\to\Xi_c^{\prime0}$}&$ f_1$&$-0.202\pm0.006\pm0.021\pm0.005$&0.89&5.69&-0.21&-0.38\cr\cline{2-7}
&$ f_2$&$0.069\pm0.058\pm0.046\pm0.012$&407.74&14.72&0.07&0.16\cr\cline{2-7}
&$ f_3$&$-0.065\pm0.049\pm0.040\pm0.016$&8.76&5.84&-0.07&0.27\cr\cline{2-7}
&$ f_4$&$0.433\pm0.035\pm0.056\pm0.010$&0.87&4.53&0.50&-3.10\tabularnewline
\hline\hline
\multirow{4}{*}{$\Omega_b^-\to\Omega_c^0$}&$ f_1$&$-0.190\pm0.005\pm0.021\pm0.006$&0.78&5.27&-0.203&0.67\cr\cline{2-7}
&$ f_2$&$0.074\pm0.061\pm0.048\pm0.015$&19.08&7.05&0.077&-0.13
\cr\cline{2-7}
&$ f_3$&$-0.066\pm0.052\pm0.042\pm0.019$&4.19&4.87&-0.076&0.521\cr\cline{2-7}
&$ f_4$&$0.427\pm0.035\pm0.061\pm0.015$&0.78&4.30&0.504&-3.88\tabularnewline
\hline\hline
\end{tabular}
\end{table}

\end{widetext}

For  extrapolating the form factors to the full $q^2$ region,  we use the Bourrely-Caprini-Lellouch (BCL) parametrization~\cite{Boyd:1997kz,Caprini:1997mu,Bourrely:2008za,Bharucha:2010im} in which the form factors are expanded in powers of a conformal mapping variable. The BCL parametrization is shown as
 \begin{eqnarray}
f(t)&=&\frac{1}{1-t/m_R^2}\sum^{k_{max}}_{k=0} \alpha_kz^k(t,t_0),\nonumber\\
z(t,t_0)&=&\frac{\sqrt{t_+-t}-\sqrt{t_+-t_0}}{\sqrt{t_+-t}+\sqrt{t_+-t_0}},\notag\\ t_0&=&t_+\big(1-\sqrt{1-\frac{t_-}{t_+}}\big),\nonumber\\ 
t_\pm&=&(m_{{\cal B}_b}\pm m_{{\cal B}_c})^2.
\label{BCL}
\end{eqnarray}
The $m_R$ are the masses of the low-laying $B_c$ resonance.  Additionally, the $q^2$ dependence of form factors can also be described by the pole model: 
 \begin{eqnarray}
	F(q^2)=\frac{F(0)}{1-\frac{q^2}{m_{\rm fit}^2}+\delta(\frac{q^2}{m_{\rm fit}^2})^2},\label{fit}
\end{eqnarray}
where $F(0)$ is the numerical results of form factor at $q^2=0$. Using this formula, one can fit the two parameters $m_{\rm fit}$ and $\delta$ by the numerical results of form factors with different $q^2$. When the fitting results of $m_{\rm fit}$ are imaginary results, we need to modify the parametrization scheme as
 \begin{eqnarray}
	F(q^2)=\frac{F(0)}{1+\frac{q^2}{m_{\rm fit}^2}+\delta(\frac{q^2}{m_{\rm fit}^2})^2}.\label{fit1}
\end{eqnarray}
In this work, we analyze the $q^2$ dependence of the form factors using both models in Eq.\eqref{BCL}, Eq.\eqref{fit}, and Eq.\eqref{fit1}, respectively. Our strategy is to calculate the form factors at $q^2=\{0,-0.1,-0.5,-0.7,-1,-1.5\}{\rm GeV}^2$ and fit the parameters in Eq.\eqref{BCL}, Eq.\eqref{fit}, and Eq.\eqref{fit1}. Then we extend the form factors to the physical region with the fitted parameters in these two models. Table~\ref{ffnr3} shows the form factors with $q^2=0$ and the corresponding fit parameters.

\begin{widetext}

 \begin{table}[!http]
\caption{
Numerical results for the SM form factors of b-baryon to c-baryon transitions induced by $b\to c$ decays are presented. The parameters $\delta$ and $m_{\rm fit}$ are the results of fitting the pole model in Eq.~\eqref{fit}, and the $^*$ indicates that $m_{\rm fit}$ is fitted using Eq.~\eqref{fit1}.  For the form factors with $q^2=0$, i.e., $F(0)$, we estimated their uncertainties caused by the parameters in LFQM, namely, $\beta_{b[qq]}$, $\beta_{c[qq]}$, and $m_{di}$, which are varied by $10\%$.}
\label{table2} %
\begin{tabular}{|c|c|c|c|c|c|c|c|c|}
\hline \hline
\multirow{2}{*}{channel}&\multirow{2}{*}{form factor}&\multicolumn{3}{c|}{Pole model parameter}&\multirow{2}{*}{form factor}&\multicolumn{3}{c|}{Pole model parameter} \cr\cline{3-5}\cline{7-9}
&& F(0) &$\delta$ & $m_{\rm fit}$ &  &  F(0) &$\delta$ & $m_{\rm fit}$  \\
\hline
\multirow{4}{*}{$\Lambda_b^0\to\Lambda_c^+$}&$ F_1$&$0.637\pm0.026\pm0.036\pm0.008$&0.72&5.50&$G_1$&$0.625\pm0.027\pm0.022\pm0.014$&0.05&5.13\cr\cline{2-9}
&$F_2$&$-0.134\pm0.015\pm0.001\pm0.000$&0.26&3.35&$G_2$&$0.005\pm0.010\pm0.040\pm0.017$&0.48&1.18*\cr\cline{2-9}
&$F_3$&$0.026\pm0.015\pm0.010\pm0.008$&0.51&2.27&$G_3$&$-0.112\pm0.050\pm0.118\pm0.040$&2.53&3.22*\tabularnewline
\hline\hline
\multirow{4}{*}{$\Sigma_b^+\to\Sigma_c^{++}$}&$ F_1$&$0.549\pm0.020\pm0.029\pm0.003$&0.40&4.00&$G_1$&$-0.197\pm0.009\pm0.007\pm0.005$&0.07&4.60\cr\cline{2-9}
&$ F_2$&$0.551\pm0.011\pm0.021\pm0.010$&0.32&4.06&$G_2$&$0.040\pm0.006\pm0.010\pm0.009$&0.11&2.86\cr\cline{2-9}
&$ F_3$&$-0.269\pm0.007\pm0.005\pm0.011$&0.31&3.99&$G_3$&$-0.060\pm0.023\pm0.032\pm0.022$&0.37&2.27\cr\cline{2-9}
\hline\hline
\multirow{4}{*}{$\Sigma_b^-\to\Sigma_c^0$}&$ F_1$&$0.548\pm0.017\pm0.026\pm0.006$&0.41&4.00&$G_1$&$-0.196\pm0.006\pm0.004\pm0.009$&0.09&4.63\cr\cline{2-9}
&$ F_2$&$0.550\pm0.006\pm0.014\pm0.018$&0.32&4.06&$G_2$&$0.038\pm0.006\pm0.018\pm0.018$&0.10&2.90\cr\cline{2-9}
&$ F_3$&$-0.268\pm0.014\pm0.010\pm0.020$&0.31&4.00&$G_3$&$-0.057\pm0.023\pm0.051\pm0.043$&0.37&2.28\tabularnewline
\hline\hline
\multirow{4}{*}{$\Sigma_b^0\to\Sigma_c^+$}&$ F_1$&$0.548\pm0.020\pm0.029\pm0.003$&0.41&4.00&$G_1$&$-0.196\pm0.008\pm0.007\pm0.004$&0.09&4.63\cr\cline{2-9}
&$ F_2$&$0.551\pm0.012\pm0.022\pm0.011$&0.32&4.06&$G_2$&$0.039\pm0.006\pm0.011\pm0.009$&0.10&2.90\cr\cline{2-9}
&$ F_3$&$-0.268\pm0.007\pm0.005\pm0.011$&0.31&4.00&$G_3$&$-0.057\pm0.023\pm0.032\pm0.023$&0.37&2.28\tabularnewline
\hline\hline
\multirow{4}{*}{$\Xi_b^0\to\Xi_c^+$}&$ F_1$&$0.635\pm0.030\pm0.036\pm0.006$&0.75&5.56&$G_1$&$0.620\pm0.031\pm0.020\pm0.014$&0.03&5.23\cr\cline{2-9}
&$ F_2$&$-0.155\pm0.019\pm0.001\pm0.001$&0.27&3.45&$G_2$&$0.012\pm0.011\pm0.046\pm0.021$&0.09&2.10*\cr\cline{2-9}
&$ F_3$&$0.028\pm0.017\pm0.13\pm0.011$&0.49&2.40&$G_3$&$-0.154\pm0.061\pm0.141\pm0.054$&4.56&4.05*\tabularnewline
\hline\hline
\multirow{4}{*}{$\Xi_b^-\to\Xi_c^0$}&$ F_1$&$0.633\pm0.030\pm0.036\pm0.006$&0.74&5.55&$G_1$&$0.618\pm0.031\pm0.020\pm0.014$&0.01&5.21\cr\cline{2-9}
&$ F_2$&$-0.154\pm0.019\pm0.001\pm0.001$&0.27&3.45&$G_2$&$0.013\pm0.011\pm0.046\pm0.021$&0.96&1.86*\cr\cline{2-9}
&$ F_3$&$0.028\pm0.017\pm0.013\pm0.011$&0.50&2.29&$G_3$&$-0.157\pm0.061\pm0.141\pm0.054$&3.84&3.79\tabularnewline
\hline\hline
\multirow{4}{*}{$\Xi_b^{\prime 0}\to\Xi_c^{\prime+}$}&$ F_1$&$0.565\pm0.019\pm0.023\pm0.001$&0.44&4.11&$G_1$&$-0.197\pm0.010\pm0.006\pm0.004$&0.13&4.84\cr\cline{2-9}
&$ F_2$&$0.555\pm0.007\pm0.012\pm0.010$&0.38&4.20&$G_2$&$0.026\pm0.006\pm0.012\pm0.010$&-0.01&2.89\cr\cline{2-9}
&$ F_3$&$-0.252\pm0.008\pm0.008\pm0.013$&0.46&4.29&$G_3$&$-0.022\pm0.025\pm0.040\pm0.025$&1.04&1.71*\tabularnewline
\hline\hline
\multirow{4}{*}{$\Xi_b^{\prime -}\to\Xi_c^{\prime0}$}&$ F_1$&$0.565\pm0.019\pm0.023\pm0.001$&0.44&4.11&$G_1$&$-0.197\pm0.010\pm0.006\pm0.004$&0.13&4.84\cr\cline{2-9}
&$ F_2$&$0.555\pm0.007\pm0.012\pm0.010$&0.38&4.20&$G_2$&$0.025\pm0.006\pm0.012\pm0.010$&-0.01&2.89\cr\cline{2-9}
&$ F_3$&$-0.252\pm0.008\pm0.008\pm0.013$&0.46&4.29&$G_3$&$-0.021\pm0.025\pm0.040\pm0.025$&1.04&1.71*\tabularnewline
\hline\hline
\multirow{4}{*}{$\Omega_b^-\to\Omega_c^0$}&$ F_1$&$0.532\pm0.021\pm0.026\pm0.004$&0.44&3.95&$G_1$&$-0.184\pm0.011\pm0.007\pm0.006$&0.17&4.60\cr\cline{2-9}
&$ F_2$&$0.529\pm0.010\pm0.015\pm0.016$&0.37&4.00&$G_2$&$0.021\pm0.007\pm0.011\pm0.011$&0.11&2.58\cr\cline{2-9}
&$ F_3$&$-0.245\pm0.005\pm0.017\pm0.019$&0.40&4.00&$G_3$&$-0.009\pm0.026\pm0.041\pm0.028$&-0.07&0.26\tabularnewline
\hline
\end{tabular}
\end{table}
\end{widetext}

\begin{widetext}

\begin{figure}[htbp!]
	\begin{minipage}[t]{0.45\linewidth}
		\centering
		\includegraphics[width=1\columnwidth]{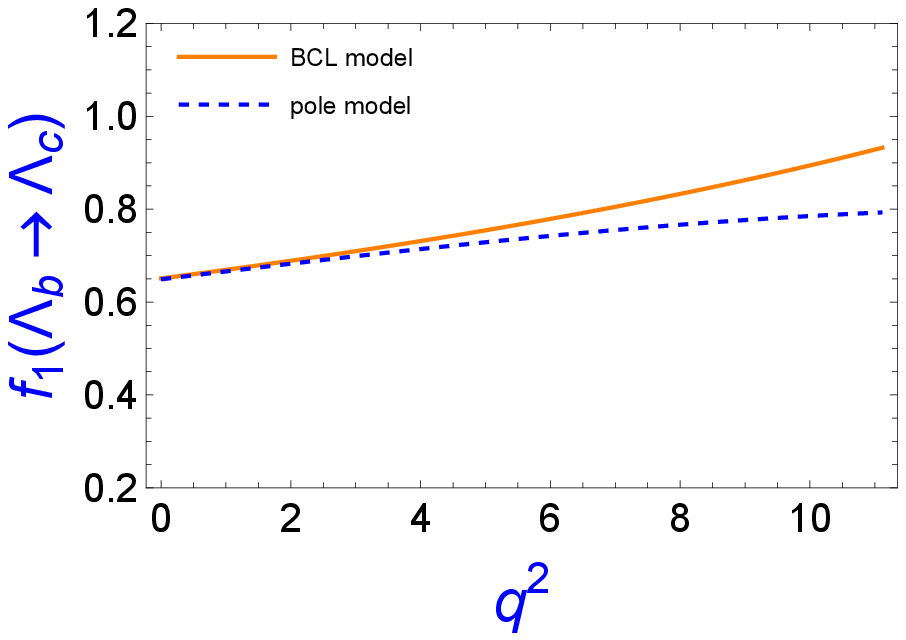}
	\end{minipage}
	\begin{minipage}[t]{0.45\linewidth}
		\centering		\includegraphics[width=1\columnwidth]{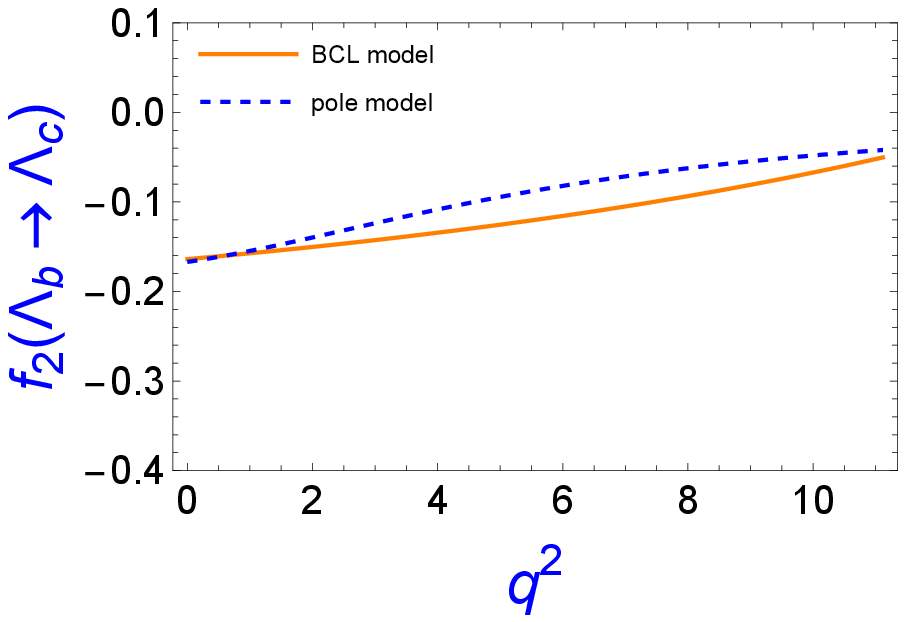}
	\end{minipage}
	\begin{minipage}[t]{0.45\linewidth}
		\centering
		\includegraphics[width=1\columnwidth]{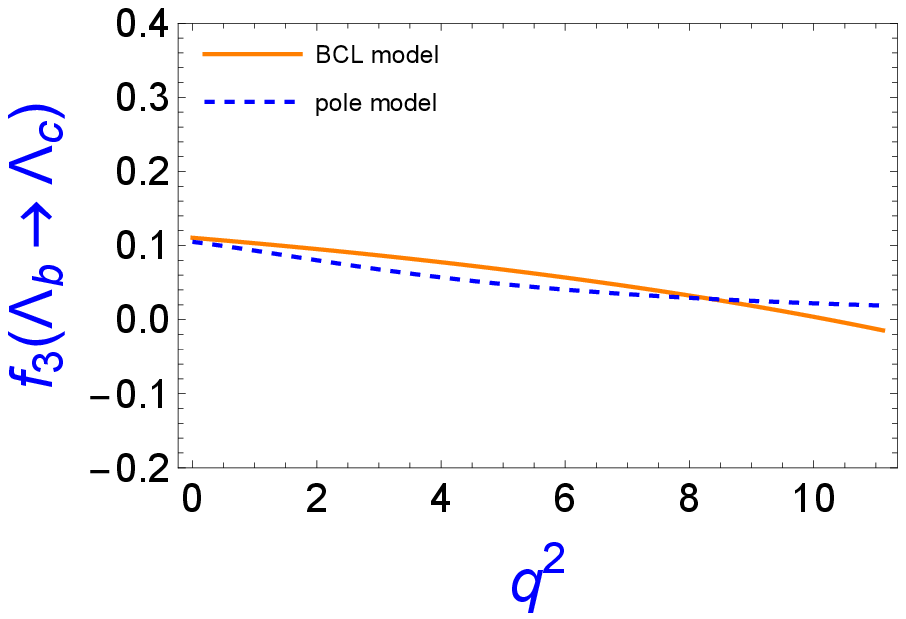}
	\end{minipage}
	\begin{minipage}[t]{0.45\linewidth}
		\centering
		\includegraphics[width=1\columnwidth]{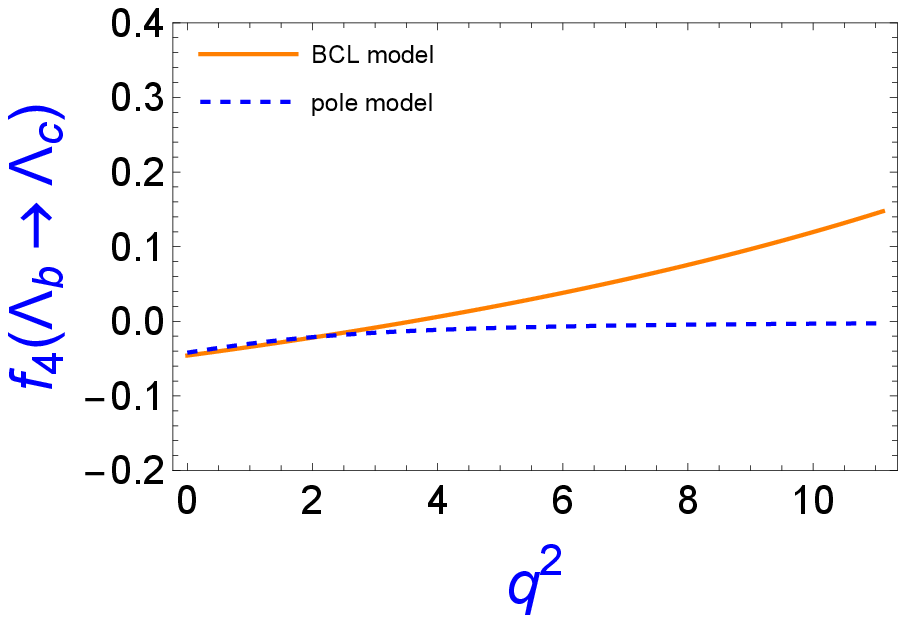}
	\end{minipage}
	\caption{ The $q^2$ dependence form factors of $\Lambda_b\to\Lambda_c$ process with BCL model (Bule line) and pole model (Orange line.)}
\label{fig:1}
\end{figure} 

\end{widetext}

Except the  tensor form factor, the form factors defined by the vector and axial-vector current are also estimated for completeness, which are defined as
  \begin{eqnarray}
&&\langle {\mathcal B}^\prime(P^\prime,S^\prime)|\bar c \gamma^\mu(1-\gamma_5)b|{\mathcal{B}(P,S)}\rangle \notag\\
&&=\bar u(P^\prime,S^\prime)\bigg[F_1(q^2)\gamma^\mu+F_2(q^2)\frac{i\sigma^{\mu\nu}q_\nu}{M_{\mathcal B}}\notag\\
&&+F_3(q^2)\frac{q^\mu}{M_{\mathcal B}}-\bigg(G_1(q^2)\gamma^\mu+G_2(q^2)\frac{i\sigma^{\mu\nu}q_\nu}{M_{\mathcal B}}\notag\\
&&+G_3(q^2)\frac{q^\mu}{M_{\mathcal B}}\bigg)\gamma_5]u(P,S).
\end{eqnarray}
Numerical results of these form factors are given in Table.~\ref{table2} and our results are generally consistent with the previous work~\cite{Ke:2019smy,Miao:2022bga}. The $q^2$ dependence of these form factors are also fitted with the pole model in Eq.~\eqref{fit} and Eq.~\eqref{fit1} since the validity of the pole model with the vecor and axial-vector form factors in LFQM has been verified in many studies~\cite{Ke:2007tg,Ke:2012wa,Ke:2017eqo,Ke:2019smy}.

\begin{widetext}


\begin{figure}[htbp!]
	\begin{minipage}[t]{0.45\linewidth}
		\centering
		\includegraphics[width=1\columnwidth]{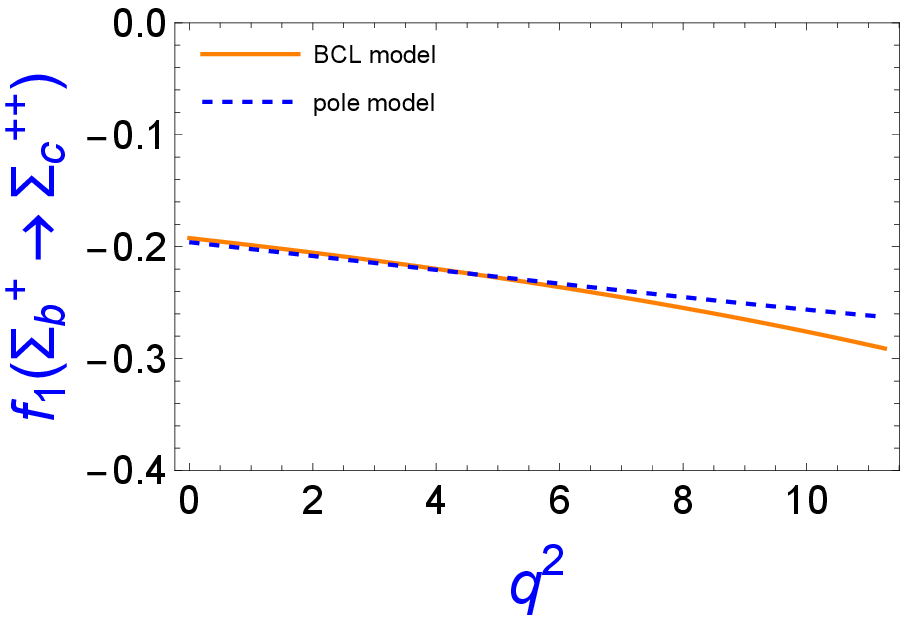}
	\end{minipage}
	\begin{minipage}[t]{0.45\linewidth}
		\centering
		\includegraphics[width=1\columnwidth]{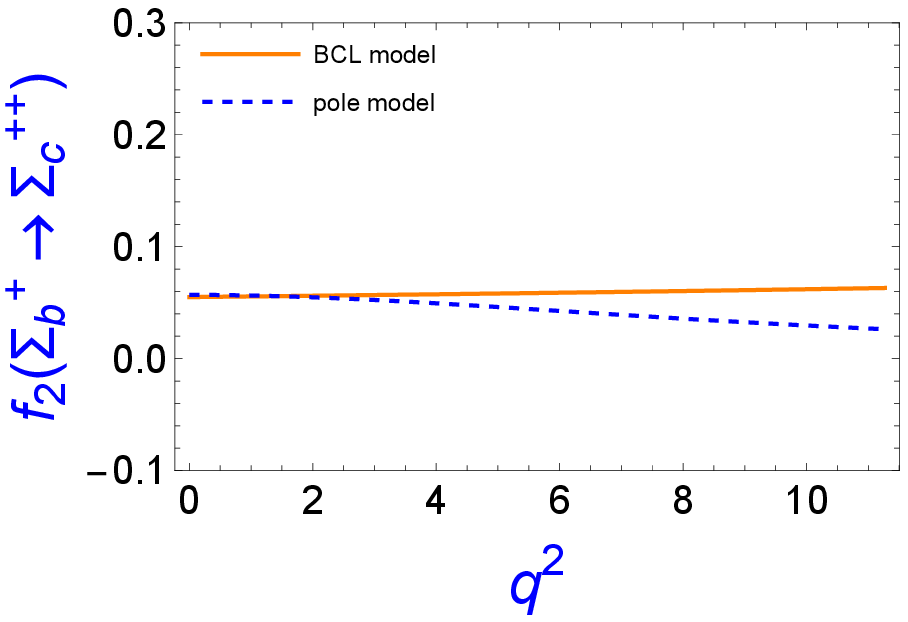}
	\end{minipage}
	\begin{minipage}[t]{0.45\linewidth}
		\centering
		\includegraphics[width=1\columnwidth]{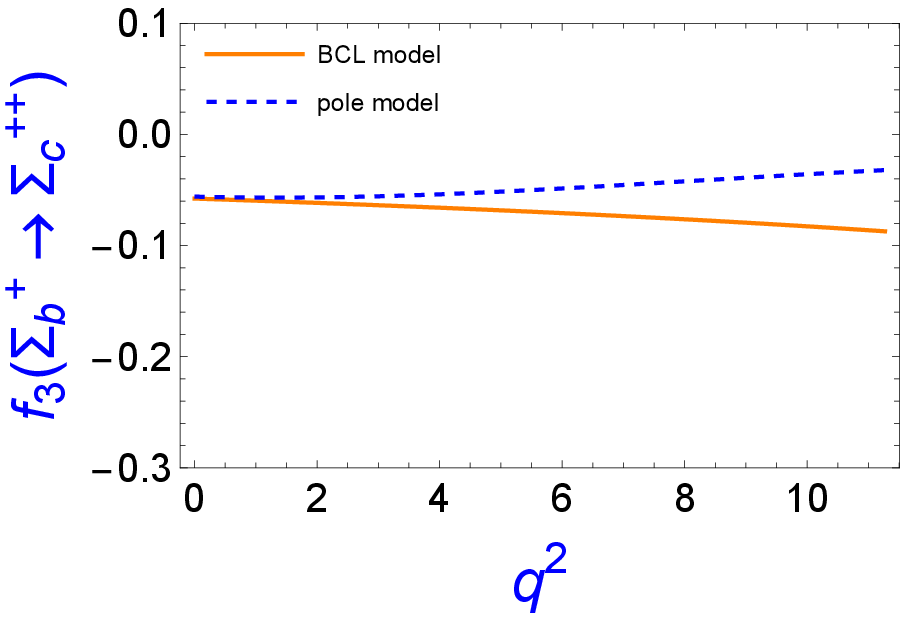}
	\end{minipage}
	\begin{minipage}[t]{0.45\linewidth}
		\centering
		\includegraphics[width=1\columnwidth]{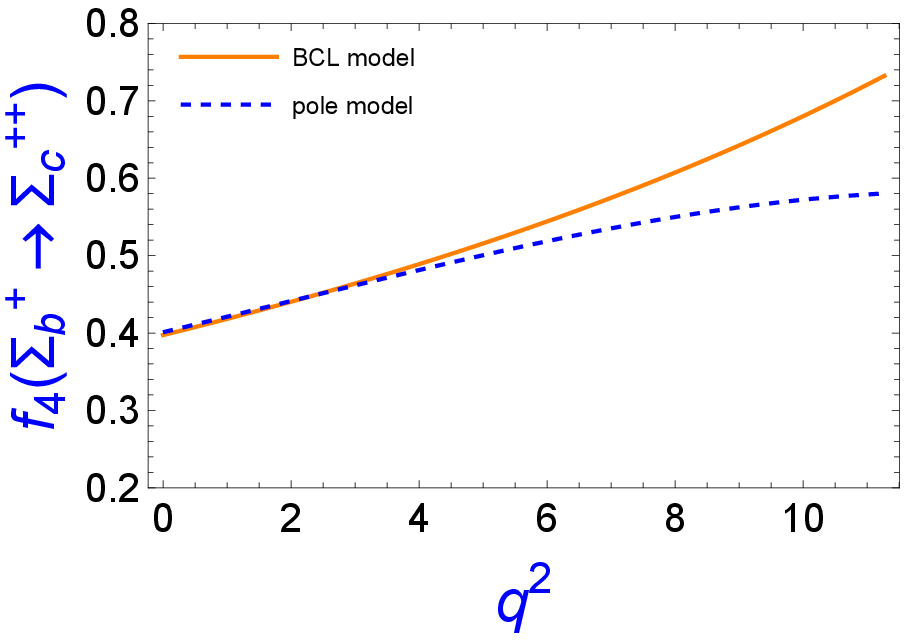}
	\end{minipage}
	\caption{ The $q^2$ dependence form factors of $\Sigma_b^+\to\Sigma_c^{++}$ process with BCL model (Bule line) and pole model (Orange line.)}
\label{fig:2}
\end{figure} 

\end{widetext}

To analyze the $q^2$ dependence of the form factors, we also plot the results of the form factors as functions of $q^2$ in Fig.~\ref{fig:1} and Fig.~\ref{fig:2}. Since the main focus is   the tensor form factors for two types of processes, $\bar 3\to \bar 3$ and $6\to 6$, we use $\Lambda_b\to\Lambda_c$ and $\Sigma^+_b\to\Sigma^{++}_c$ as examples.
From Fig.~\ref{fig:1} and Fig.~\ref{fig:2}, one can see that the fit results with two different models are broadly consistent with each other. However, the $q^2$-dependent form factor $f_4$ exhibits a large discrepancy with $q^2\sim (m_{{\cal B}_b}-m_{{\cal B}_c})^2$ for both models. In the analysis presented in Ref.~\cite{Liu:2022mxv}, the form factor has a pole structure corresponding to the specific current. In our work, the pole mass $m_{pole}$ should be set to $m_{pole}=m_{B_c}$, which is consistent with the BCL model in Eq.\eqref{BCL}. However, the fit result of $f_4$ with the pole model is different from our conclusion, especially for the $\Lambda_b\to\Lambda_c$ process. Therefore,  it is likely that the BCL model  describes the $q^2$ dependence of the form factor better. Nonetheless, we will still present the results obtained using the pole model since this model is widely used in the light-front approach analysis~\cite{Chang:2020wvs,Hu:2020mxk}.

\section{Summary}

In our work, we presented an exploration  of the  form factors in the hadronic matrix element with a tensor current. 
With the help of the LFQM, hadron states can be expressed in terms of relativistic quark-diquark configurations, from which the form factors are extracted. In our calculation, we have calculated the form factors in the $q^2< 0$ region  and accessed  their $q^2$ dependence using both the Bourrely-Caprini-Lellouch parametrization model and the pole model. The difference between the two models is shown in Fig.~\ref{fig:1} and Fig.~\ref{fig:2} and the results with two different models are generally consistent with each other. 

Our work provides the ingredients for further research on the tensor current and its contribution in NP analysis. The phenomenological  analysis of new physics contributions made on these form factors can   serve as a useful reference for future experimental and theoretical studies.

\section*{Acknowledgements}

We thank  Dr. Fei Huang and Mr. Chang Yang for useful discussions. 
This work  is supported in part by Natural Science Foundation of China under grant No. U2032102, 12090064, 12125503,  by Natural Science Foundation of Shanghai, and  by China Postdoctoral Science Foundation under Grant No. 2022M72210.


\end{document}